\documentclass{aastex631} %comment out for submission draft
\usepackage{psfig,epsfig,amsfonts,amsmath,amssymb,graphicx,morefloats,appendix,tensor}
\usepackage{color}
\usepackage{natbib}
\usepackage{hyperref}
\usepackage{subfigure}

\begin{document}

\title{TESS Discovers a Second System of Transiting Exocomets in the Extreme Debris Disk of RZ Psc}

\author[0009-0006-5310-6855]{Adalyn Gibson}
\affil{Department of Astrophysical and Planetary Sciences, University of Colorado, 2000 Colorado Avenue, Boulder, CO 80309, USA}

\author[0000-0001-7891-8143]{Meredith A. MacGregor}
\affil{Department of Physics and Astronomy, Johns Hopkins University, 3400 N Charles St, Baltimore, MD 21218, USA}

\author[0000-0002-0583-0949]{Ward S. Howard}
\affil{Department of Astrophysical and Planetary Sciences, University of Colorado, 2000 Colorado Avenue, Boulder, CO 80309, USA}
\affil{NASA Hubble Fellowship Program Sagan Fellow}

\author[0000-0002-3656-6706]{Ann Marie Cody}
\affil{SETI Institute, 339 N Bernardo Avenue Suite 200, Mountain View, CA 94043, USA}

\author[0009-0001-4487-7299]{Mark Swain}
\affil{Jet Propulsion Laboratory, California Institute of Technology, 4800 Oak Grove Drive, Pasadena, CA 91109, USA}

\author[0000-0002-0040-6815]{Jennifer A. Burt}
\affil{Jet Propulsion Laboratory, California Institute of Technology, 4800 Oak Grove Drive, Pasadena, CA 91109, USA}

\author[0000-0002-4115-0318]{Laura Venuti}
\affil{SETI Institute, 339 N Bernardo Avenue Suite 200, Mountain View, CA 94043, USA}

\author[0000-0002-7260-5821]{Evgenya Shkolnik}
\affil{School of Earth and Space Exploration, Arizona State University, Tempe, AZ 85287, USA}

\author[0000-0001-8292-1943]{Neal J. Turner}
\affil{Jet Propulsion Laboratory, California Institute of Technology, 4800 Oak Grove Drive, Pasadena, CA 91109, USA}

\author{Alan Didion}
\affil{Jet Propulsion Laboratory, California Institute of Technology, 4800 Oak Grove Drive, Pasadena, CA 91109, USA}

\author{Jaime Nastal}
\affil{Jet Propulsion Laboratory, California Institute of Technology, 4800 Oak Grove Drive, Pasadena, CA 91109, USA}

\author{David Makowski}
\affil{Jet Propulsion Laboratory, California Institute of Technology, 4800 Oak Grove Drive, Pasadena, CA 91109, USA}

\begin{abstract}
We present the TESS discovery of only the second system of transiting exocomets with a sufficient number of events to measure the size distribution in the RZ Psc system, enabling comparisons with the $\beta$ Pictoris and Solar System size distributions. Twenty-four transits with absorption depths (AD) of 1--20\% were observed across three TESS sectors of the 20-50 Myr K0V star, detected as part of our TESS survey of extreme debris disks identified by their IR excess. We discover that the ADs (and hence exocomet radii) follow a broken power-law cumulative frequency distribution not previously seen in extrasolar contexts but similar to that observed in Solar System Kuiper Belt Object sizes, with power-law slopes above and below the break of $\gamma_\mathrm{AD>break}$=2.32$\pm$0.12 and $\gamma_\mathrm{AD<break}$=0.11$\pm$0.04, respectively. We derive size distributions of 1--7~km from two independent lines of evidence. We use the RZ Psc exocomet rate to predict exocomet yields for the Early eVolution Explorer (EVE) NASA astrophysics Small Explorer (SMEX) mission concept to obtain simultaneous photometry of 10$^4$ young stars in NUV, optical, and NIR bands. Assuming occurrence rates scaled from RZ Psc, EVE would detect 590 exocomets from $\approx$70 young systems in the optical band, with $\approx$120 simultaneous 5$\sigma$ detections in all three bands. These data would enable grain sizes of 200--700~nm and graphite--olivine compositions of dozens of events to be distinguished at 2.5--3$\sigma$, as well as a 4$\sigma$ determination of the accuracy of the Herschel-derived M-debris disk fraction.
\end{abstract}


\section{Introduction}
\label{sec:intro}

Debris disks are extrasolar analogs of the Asteroid and Kuiper Belts in our Solar System.  In these remnant belts, planetesimals (similar to asteroids and comets) collide and grind down in a collisional cascade \cite[e.g.,][]{gaspar:2012}.  Typically, we detect the smallest grains ($\mathrm{\mu}$m--cm sizes) produced in this cascade through scattered light or thermal emission.  Based on the presence of these small grains, we infer the existence of a `birth ring' of larger planetesimals \cite[e.g.,][]{beust:2006,strubbe:2006}. In the $\beta$ Pictoris system, however, 1.5 to 6.7 km radii bodies have been observed directly via transits in front of the host star \citep{zieba:2019}.  Enough exocomets have been discovered in this system, in fact, that constraints can be placed on both the size distribution \citep{lecavelier:2022} and cometary orbits \citep{heller:2024}.

The asymmetric distribution of CO gas in the $\beta$ Pictoris disk suggests at least one recent large collision \citep{dent:2014}.  However, it is a relatively dynamically quiet system in comparison to the population of so-called `extreme' debris disks that exhibit high fractional dust luminosities ($f\geq10^{-2}$) and significant variability in their infrared flux presumably due to numerous massive collisions \citep{meng:2014,meng:2015,su:2019,su:2020,rieke:2021,moor:2021,moor:2022}.  The majority of these disks are found around stars $<200$~Myr old, the approximate time frame for the end of terrestrial planet accretion \citep{chambers:2013,quintana:2016}.  These properties point to major collisional events as the source of variability in extreme debris disks, which could evolve on orbital timescales (months to decades).  Several of these systems also exhibit irregular dips detected by optical surveys \citep{gaidos:2019,powell:2021,melis:2021} that are likely due to large dust clumps or planetesimal fragments transiting in front of the host star.  Thus, extreme debris disks offer an exciting opportunity to constrain the properties of the large bodies producing the collisional cascade that results in smaller dust grains.

RZ Psc has been studied extensively as a young, Sun-like \citep[K0V; $R_\star/R_\odot = 1.0 $; $M_\star/M_\odot = 1.1$; $d = 184.1 \pm 0.9 \ \text{pc}$,][]{su:2023} star that hosts an extreme debris disk and could give a unique window into the structure of inner planetary systems around the time of gas disk dispersal. The age and evolutionary status of the system have been much debated in the literature.  X-ray monitoring and high-resolution optical spectroscopy indicate an age of $30-50$~Myr \citep{punzi:2018}.  Analysis of Gaia DR2 astrometric data suggests that it could be a member of the Cas-Tau OB association, which is slightly younger at $20^{+3}_{-5}$~Myr \citep{potravnov:2019}.  The system has been monitored at optical wavelengths for several decades and exhibits frequent irregular dips (up to $\sim$ 5\% of the flux) due to stellar variability \citep{grinin:2010,dewit:2013,kennedy:2017}. The level of optical variability is characteristic of the Herbig Ae star UX Orionis (UXOr) dimming events common among pre-main sequence stars.  However, the estimated gas accretion rate is low compared to typical protoplanetary disks \citep{potravnov:2017}.  ALMA imaging at 1.3~mm reveals a compact and highly inclined dust disk spanning $\sim0.1 - 13$~au without cold gas \citep{su:2023}.  The disk is likely truncated due to a 0.12~M$_\odot$ companion at a projected separation of 22~au \citep{kennedy:2020}.  Given its location, this companion is expected to perturb the disk significantly exciting collisions that can explain both the variation in infrared flux through production of small dust and the optical dips due to transits of larger bodies.

Here, we present analysis of archival data from the Transiting Exoplanet Survey Satellite (TESS) from sectors 17, 57, and 84 targeting RZ Psc.  We detect 24 dips in the TESS light curves, which we interpret and model as falling evaporating bodies (FEBs) or `exocomets'.  This analysis yields new insights into the dynamics of this exciting system and the larger population of extreme debris disks.  In \S\ref{sec:observations} we discuss TESS observations and full-frame image data reduction. In \S\ref{sec:analysis}, we describe FEB identification and modeling of the size distribution.  In \S\ref{sec:disc}, we compare the size RZ Psc size distribution to $\beta$ Pictoris and the Solar System.  We also discuss multi-wavelength observations of exocomets with the Early eVolution Explorer (EVE) NASA Small Explorer (SMEX) mission concept and future observations with TESS, Vera C. Rubin Observatory (VRO), and the Nancy Grace Roman Space Telescope.

\section{Observations and Light Curve Generation}
\label{sec:observations}

RZ Psc (TIC 15693497) was observed by TESS in sectors 17, 57, and 84. These sectors were executed during the following times: 2019-10-08 04:15:26 -- 2019-11-02 04:37:25 (Sector 17), 2022-09-30 20:20:44 -- 2022-10-29 14:42:43 (Sector 57), and 2024-10-01 01:59:36 -- 2024-10-26 20:09:35 (Sector 84).  The total on-source science time combining all three sectors is 60.50~days. Sector 17, which took place during the TESS Prime Mission, has a cadence of 30~min. Sectors 57 and 84, which took place during the TESS Second Extended Mission, have a cadence of 200~s. Since RZ Psc was not a specific TESS target, pipeline-reduced SPOC postage stamp based light curves are not available. The \texttt{\detokenize{eleanor}}\footnote{\url{https://github.com/afeinstein20/eleanor}} package reduces TESS Full Frame Images (FFIs) into corrected light curves for analysis, in units of normalized flux. We used \texttt{\detokenize{eleanor}} to create a target pixel file around RZ Psc, and then employed a PCA-based systematics removal method to extract the light curve of RZ Psc. We used the predetermined aperture and employed 1D Target Pixel File (TPF) background subtraction to correct the background. 

We removed standard systematic noise sources in the \texttt{eleanor} light curves, including background contamination and momentum dumps, which occurred several times per sector. Background contamination was determined by comparison with nearby sources ($\leq350\arcsec$). This step also lowers the risk of false-positives due to common-time transits encountered in the \texttt{eleanor-lite} exocomet search in \citet{norazman:2025}. The time points of contamination were determined and masked from the observations used here. By determining the flux averages of the first ten time points before and after the momentum dump and calculating the difference, we were able to correct the flux following the momentum dump via the difference. We identified two flares in sector 17, with a peak flux of \textless 1\% higher than that of pre-flare flux at 1785.7 TESS Barycentric Julian Days (TBJD) and 1786.45 TBJD. Two flares were also observed during sector 84 at 3600.325 and  3600.420 TBJD, with a peak flux of 1.7\% and 0.4\% higher than that of pre-flare flux respectively. We further discuss decontamination of stellar variability in Appendix~\ref{GP_model}. Figure~\ref{fig:visit_LC} shows the three sectors in their entirety with the stellar variability subtracted.

\begin{figure}
    \centering
    \includegraphics[width=\textwidth]{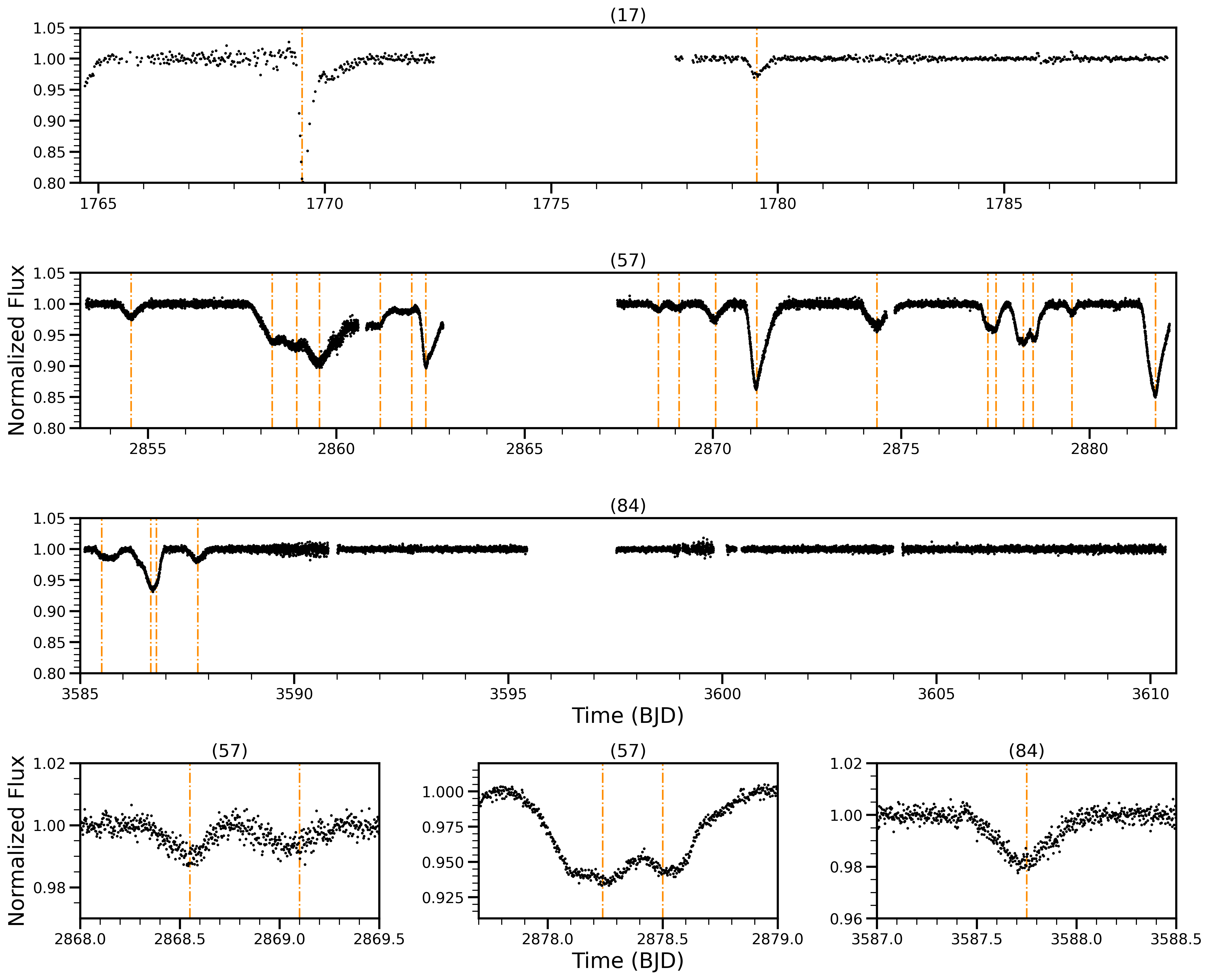}
    \caption{The top three panels show all three TESS sectors from start to finish, with the stellar variability and outliers subtracted. Each exocomet is denoted by the dashed line (orange). The bottom three panels are zoomed in examples in sectors 57 and 84.}
    \label{fig:visit_LC}
\end{figure}

\section{Results and Analysis: The Exocomets of Rz Psc}
\label{sec:analysis}

We present an in-depth analysis of exocomets in the RZ Psc system.  This is only the second exocomet host system studied in this way following the $\beta$ Pictoris system \citep{zieba:2019, lecavelier:2022, heller:2024}.

\subsection{Exocomet Identification}\label{sec:identification}
We first identified exocomets visually before modeling their light curves as described in Section~\ref{sec:modeling}. Exocomets have an identifiable asymmetric transit shape with a sharp ingress followed by a slowly decaying egress phase \citep{Kennedy:2019, Lecavelieretal:1999}. The sharp initial drop in brightness is caused by optically thick dust transiting the star. The slowly decaying egress is due to the transit of the comparatively optically thin exocomet tail. In all three sectors of TESS observations, we identified 24 exocomet transits, which implies a transit rate of 0.40 exocomets day\textsuperscript{-1}. Many of these were part of multi-cometary transit events, with only 8 single comet transits identified. \cite{Lecaveliel:1999} and \cite{Lecavelieretal:1999} predict that exocomet signals would dim tenths of a percent of the flux, which we confirm here.  We do note that $\beta$ Pictoris and RZ Psc are significantly different spectral type stars; the same event that would produce a dip of $0.1\%$ for $\beta$ Pictoris would result in a $0.3\%$ dip for RZ Psc. RZ Psc being an extreme debris disk also contributes to the difference in depth from the $\beta$ Pictoris system. In addition, 5 transits have a significantly larger depth $\geq$ 10\% of the baseline. We interpret the RZ Psc events as exocomets due to the similarity in observed light curve morphologies and transit durations to the $\beta$ Pictoris and \citet{norazman:2025} exocomet samples ($\lessapprox$1~day), rather than the broader range of morphologies and timescales ($\approx$1--10~day) usually observed from dipper stars, for pre-main sequence dipper stars, dipping depths tend to be much larger than 10\% (e.g., \citealt{boyajian:2018, capistrant:2022, tzanidakis:2025}). We note substantial overlap exists between the populations of dipper and exocomet host stars, as some dipper behavior is attributed to exocomet swarms and dust disk occultations \citep{boyajian:2018, tzanidakis:2025}. However, similarly short-duration exocomet transits are usually detected around A to G-type stars, while low-mass stars dominate the broader transit duration sample of dippers \citep{capistrant:2022}.

\subsection{Exocomet Modeling}\label{sec:modeling}

\begin{figure}
    \centering
    \includegraphics[width=\textwidth]{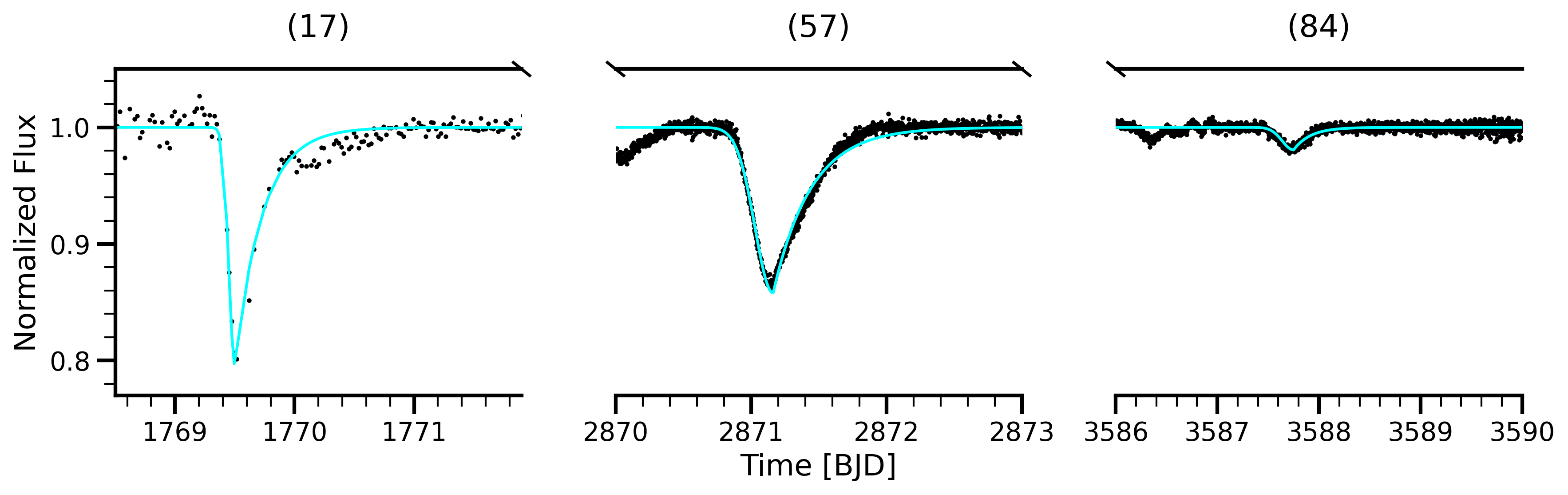}
    \caption{Examples of exocomet models across all three sectors, in the left panel sector 17 is shown with the highest AD ($AD\simeq0.20$, $R = 5.27 \pm 0.069$~km) exocomet modeled, the middle panel shows an ($AD\simeq0.14$, $R = 5.086 \pm 0.13$~km) exocomet in sector 57, and the right panel shows a smaller exocomet ($AD\simeq0.019$, $R = 1.16 \pm 0.14$~km) modeled in sector 84.}
    \label{fig:Exo_model}
\end{figure}

To model the characteristic shape of exocomet transits, we adopt a model similar to the one \cite{zieba:2019} and \cite{lecavelier:2022} applied to the $\beta$ Pictoris light curves.  The sharp initial ingress is modeled using a 1-D Gaussian function. The egress is modeled using an exponential decay function to represent the concave-down return to baseline. We define the single-component exocomet transit model as follows:

\begin{equation}\label{eqn:1}
    \mathcal{F}(t) = \begin{cases}
       1 + G(t) & t < t_c \\
      1 - A \times (b^{-c(t-t_c)}) & t \geq t_c
    \end{cases}
\end{equation}

\noindent where $G(t)$ is the 1-D Gaussian function with an amplitude of $A$, a mean of $t_c$, and a standard deviation of $\sigma$, implemented with \texttt{astropy.modeling.models.Gaussian1D}. The remaining variables are $b$ (the base of the exponential function controlling the egress), $c$ (the exponent controlling the egress steepness), and $\mathcal{F}(t)$ (the model flux).  We use a non-linear least squares model optimization implemented with \texttt{scipy.optimize.curve\_fit} to fit all variables. In order to fit multiple-component (i.e., more than one exocomet) transits, we fit summations of the single-component model described by Equation~\ref{eqn:1}, in cases where a sufficient time gap ($\sim$ 0.1 days) exists between exocomet modeling components, we identify each separate component as an exocomet transit. Examples of single exocomet transits and their best-fit single-component model are shown in Figure~\ref{fig:Exo_model}.

\subsection{Exocomet Size Distribution}\label{sec:sizedist}

\begin{figure}
    \centering
    \includegraphics[width=0.7\textwidth]{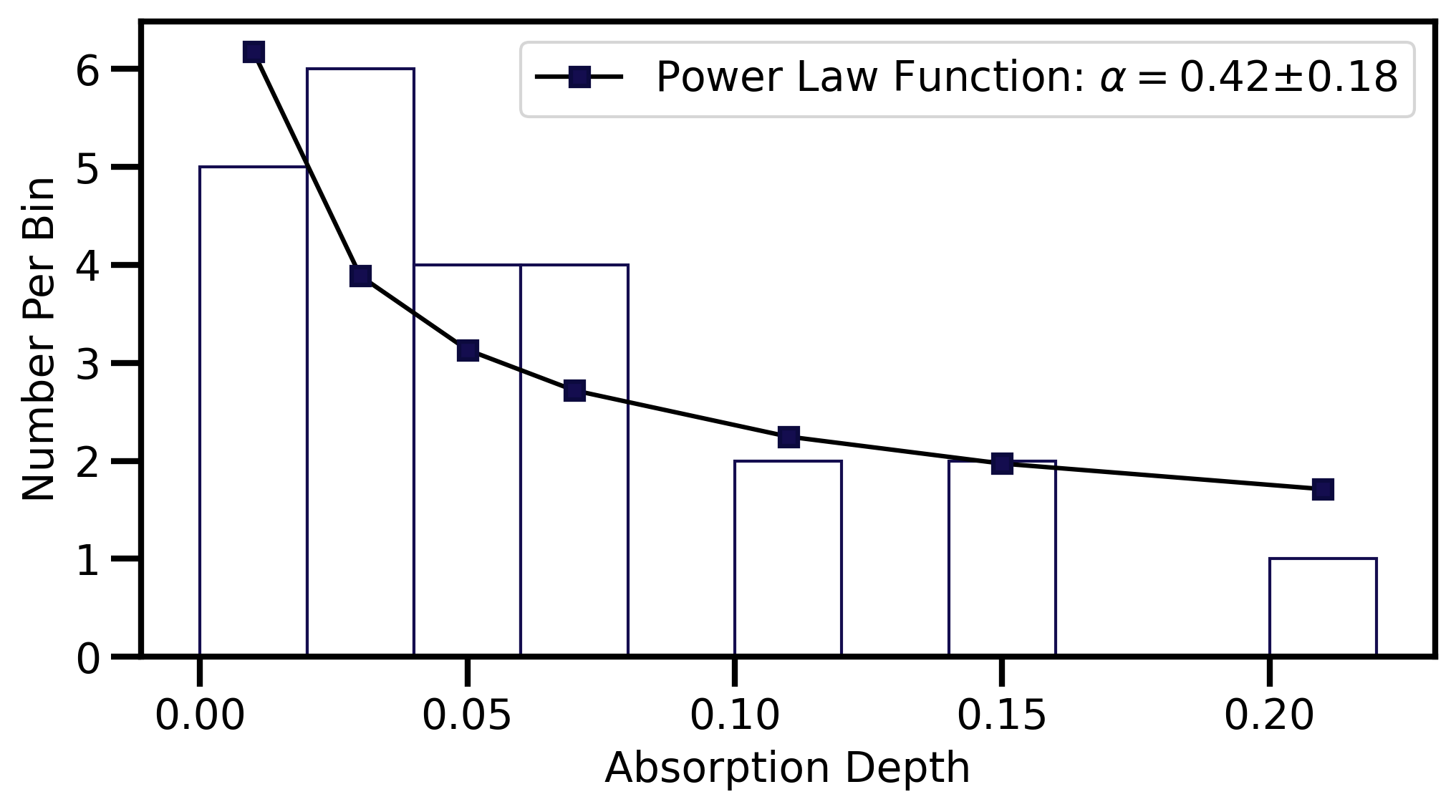}
    \caption{Non-cumulative histogram of the number of exocomets as a function of $AD$, the bin width is 0.02. We report a power law index $\alpha = 0.42 \pm 0.18$, which was calculated from Equation~\ref{eqn:2}.}
    \label{fig:ExocometHist}
\end{figure}

For each modeled exocomet, we define the absorption depth, $AD$, as the exocomet model baseline minus the maximum model dip.  We then construct a histogram of the number of exocomets as a function of absorption depth shown in Figure~\ref{fig:ExocometHist}. The shape of the histogram can be described by a power law of the form:

\begin{equation}\label{eqn:2}
    \mathcal{N} = \mathcal{A} \times AD^{-\alpha}
\end{equation}

\noindent where $\mathcal{A}$ is a normalization factor, $\alpha$ is the power law index, and $\mathcal{N}$ is the number of exocomets. We use the non-linear least squares model optimization implemented with \texttt{scipy.optimize.curve\_fit} to fit each variable, and return a best-fit power law index $\alpha = 0.42 \pm 0.18$.

\begin{figure}[t]
    \centering
    \includegraphics[width=\textwidth]{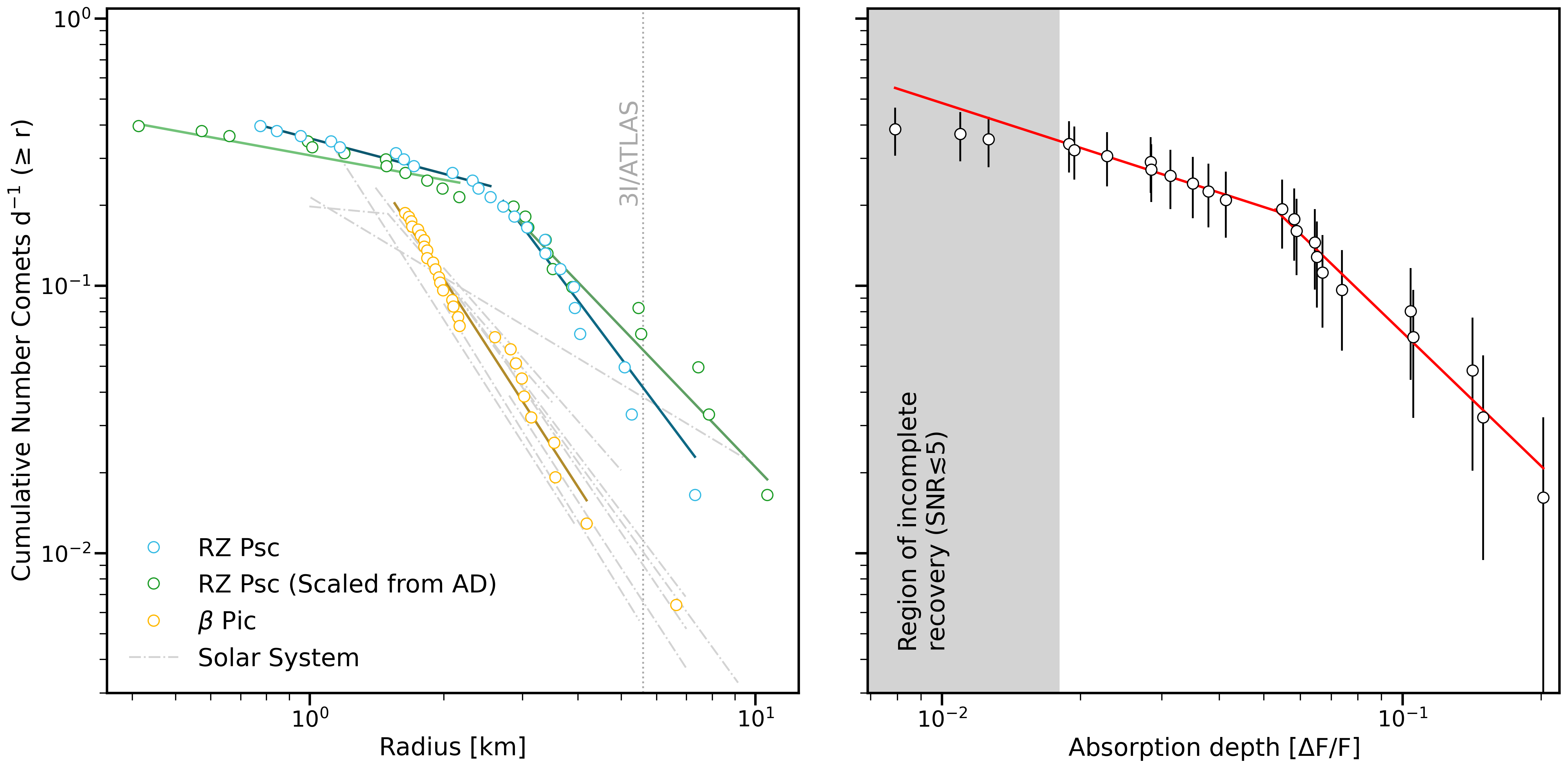}
    \caption{Left panel: (blue) Cumulative size distribution of the 24 exocomets identified in the RZ Psc system with radii in units of km, calculated from the Equations~\ref{eqn:AD}, \ref{eqn:q}, and \ref{eqn:Mdot} in Appendix~\ref{exocomet_radii_calculation}. We report a power law index $\gamma_\mathrm{R<break} = -0.12 \pm 0.02$ for the power law fit before the turnover point at $\sim$2.5~km, and a power law index $\gamma_\mathrm{R>break} = 3.45 \pm 0.17$ for the power law fit after the turnover point, which was calculated from Equation~\ref{eqn:powerlaw2}. (green) Cumulative size distribution of the 24 exocomets identified in the RZ Psc system with radii in units of km, calculated from the Equation~\ref{eqn:AD_to_rad} in Appendix~\ref{exocomet_radii_calculation}. We report a power law index $\gamma_\mathrm{R<break} = -0.39 \pm 0.03$ for the power law fit before the turnover point at around $\sim$2.26~km, and a power law index $\gamma_\mathrm{R>break} = 2.48 \pm 0.05$ for the power law fit after the turnover point. (yellow) We compare to the size distribution of $\beta$ Pictoris from \cite{lecavelier:2022}. (grey, dashed) We also compare it to solar system populations from \cite{lecavelier:2022}. (grey, dotted) We plot a vertical line represented the estimated radius of the most recent interstellar object, 3I/ATLAS, of $5.6 \pm0.7$~km from \cite{Chandler:2025}. Right panel: A comet frequency distribution in AD space to highlight our detection ability for smaller ADs. We report a power law fit: $\gamma_\mathrm{AD>break}$=2.32$\pm$0.12 and $\gamma_\mathrm{AD<break}$=0.11$\pm$0.04, where the ADs in the region of incomplete recovery were not fit. We expect that far fewer than 100\% of exocomets will be detectable in the region of incomplete recovery. This calculation was performed with a cadence of 200~s, matching the cadence of sectors 57 and 84. }
    \label{fig:ExocometSizeDistComp}
\end{figure}

We assess the completeness of our exocomet recovery by examining the cumulative exocomet $AD$ distribution in the right panel of Figure~\ref{fig:ExocometSizeDistComp}. The mean successive difference statistic (MSD; \citealt{kamat:1953}) or mean of the point-to-point photometric variation is measured for each sector in order to determine the effective noise of the light curve without regard to slowly-varying rotational trends. The MSD values range from 0.0027--0.0036, so we adopt MSD=0.0036 for a 5$\sigma$ detection threshold of $AD=0.018$. This value also corresponds to depths above which transits become readily visible by eye in visible inspection of the light curves, so we define $AD\leq0.018$ as the region of incomplete recovery (indicated by the shaded region). The majority of detected exocomet transit $AD$ values are above this cutoff, which suggests that any structure in the exocomet size distribution is likely physical and not a detection artifact. The $AD$ rates are measured as the cumulative number of comets of a given $AD$ or larger per total observation time, with rate uncertainties assuming Poisson statistics. We use Eq.\ \ref{eqn:brokenpowerAD} to fit cumulative frequency distributions to the measured values above and below the $AD$ break at 0.053, with best-fit power-law slopes of $\gamma_{AD>\mathrm{break}}$=2.32$\pm$0.12 and $\gamma_{AD<\mathrm{break}}$=0.11$\pm$0.04.
% and intercepts of $\beta_\mathrm{AD>break}$=-2.71$\pm$0.12 and $\beta_\mathrm{AD<break}$=-1.47$\pm$0.09.

Using equations adapted from \cite{lecavelier:2022} (see Equations~\ref{eqn:AD} to \ref{eqn:Mdot} in Appendix~\ref{exocomet_radii_calculation} for more detail), we calculate the radii for all detected excomets from their measured $AD$ and transit time. Figure~\ref{fig:ExocometSizeDistComp} shows the resulting cumulative size distribution for the RZ Psc exocomets (blue). The smallest detected comet has a radius of $0.77 \pm0.17$~km, while the largest has a radius $7.31 \pm 0.33$~km. The overall distribution is best described by a broken power law, with the break occurring at approximately 2.5~km. We also include a size distribution calculated via an independent technique with Equation~\ref{eqn:AD_to_rad} in Appendix~\ref{exocomet_radii_calculation} using the measured $AD$ (green). For this second method, the smallest detected comet has a radius of $0.43 \pm0.29$~km, while the largest has a radius $11.15 \pm 0.25$~km. The overall distribution is again best described by a broken power law, with the break located at approximately 2.26~km. With both of the size distributions, we fit two power laws (one below and one above the break) to the cumulative size distribution, both of the form:

\begin{equation}\label{eqn:powerlaw2}
    \mathcal{C} = \mathcal{A} \times R^{-\alpha}
\end{equation}

$\gamma$ is calculated from $\alpha$ as follows

\begin{equation}\label{eqn:gamma}
    \gamma = 2 {\alpha}-1
\end{equation}

\noindent where $R$ is the exocomet radius, $\mathcal{A}$ is a normalization factor, $\alpha$ is the power law index, and $\mathcal{C}$ is the cumulative number of exocomets. Again, we use the non-linear least squares model optimization implemented with \texttt{scipy.optimize.curve\_fit} to perform the fit and return power law indices.

\section{Discussion}
\label{sec:disc}

We have identified and modeled 24 exocomets in TESS observations of RZ Psc.  Here, we discuss the size distribution of those exocomets and how it compares with both the $\beta$ Pictoris system and our own Solar System.  We also describe the contribution of the EVE concept to the study of exocomets in young planetary systems and what other future observatories might provide.

\subsection{Size Distribution Comparison}\label{sec:sizedistcomp}

In Figure~\ref{fig:ExocometSizeDistComp}, we include cumulative size distributions measured for Solar System bodies \cite{Alvarez:2006,Bauer:2017,Boe:2019,Tancredi:2006,Snodgrass:2011,Fern:2013} and the $\beta$ Pictoris system \citep{lecavelier:2022} for comparison.  One notable difference is that more small bodies are detected in the RZ Psc system.  No exocomets are identified with radii under 1~km in the $\beta$ Pictoris system.  However, with a longer cadence, detectability of smaller exocomets is potentially reduced since their transit light curves are not as well-resolved.  All of the smaller exocomets $\lesssim1$~km in the RZ Psc system were identified in TESS sectors 57 and 84, which had a cadence of 200~s.  $\beta$ Pictoris was observed in TESS sectors 4--7 and 31--34 with cadences of 30 and 10~min, respectively. 

Longwards of the $\sim$2~km break in the size distribution, the power-law slope we derive for RZ Psc is in fairly good agreement with previous results ($\gamma_\mathrm{R>break} = 3.45\pm0.17$ and $\gamma_\mathrm{R>break} = 2.48\pm0.05$ for the two methods, respectively).  For solar system bodies, including both comets and asteroids, $\gamma$ ranges from 2.0 to 3.7.  \cite{lecavelier:2022} determine a best-fit slope for $\beta$ Pictoris of $\gamma = 3.6\pm0.8$.  In debris disks, smaller bodies are generated through a catastrophic collisional cascade where smaller ‘bullets’ shatter larger ‘targets’ through collisions \citep{Pan:2005,Pan:2012}. The classical \cite{Dohnanyi:1969} model yields $\gamma = 3.5$ for collisions where bodies are dominated by material strength and have an isotropic velocity dispersion.  \cite{Pan:2012} account for viscous stirring, dynamical friction, and collisional damping to introduce a size-dependent velocity distribution that steepens the slope to $\gamma = 3.6 - 4.0 $.  Collisions between strengthless rubble piles (i.e., bodies that are not strength-dominated) can flatten the size distribution with $\gamma = 3.0 - 3.26 $.  Observations of debris disk size distributions also agree with these values with an average value of $3.36\pm0.02$ \citep{MacGregor:2016}.  It is notable that, while the slope we measure for RZ Psc is in agreement with these other systems, the cumulative number of bodies in all size bins is larger.  This is perhaps not surprising for such a dynamically active system undergoing a larger rate of destructive collisions. 

None of the comparison size distributions extend below $\sim1$~km.  It is difficult to detect small solar system bodies, and surveys have in fact reported a deficit of such bodies \cite[e.g.,][]{Bernstein:2004}.  Theoretical work explores this parameter space more fully. \cite{Pan:2005} modify the \cite{Dohnanyi:1969} model to account for rubble piles with low material strength by requiring the kinetic energy of the bullet to equal the potential energy of the target.  This predicts a break in the size distribution at $r_\text{break}$, the size of the largest body to experience a destructive collision in a given time interval.  This break moves from smaller to larger sizes as the system ages.  At the age of the Solar System, $r_\text{break}$ is predicted to be $\sim20-50$~km.  At 0.2~Gyr, the break moves downwards in size to $\sim10$~km.  Since RZ Psc is only $\sim20-50$~Myr, our observed break of $\sim$2~km is reasonable. \cite{Schlichting:2013} explore the size distribution of Kuiper Belt Objects (KBOs) with radii of $0.01-30$~km and show a more complicated structure to the size distribution with multiple breaks and slope changes. We do note that the best-fit slope for RZ Psc below the break ($\gamma_\mathrm{R<break} = -0.12 \pm 0.02$ and $\gamma_\mathrm{R<break} = -0.39 \pm 0.03$ for the two methods, respectably) is shallower than the model predictions.  This could result from the fact that we lose detection completeness for the smallest bodies (see Figure~\ref{fig:ExocometSizeDistComp}).

There are now three currently known interstellar objects \citep{Seligman:2025}. 3I/ATLAS is largest of these objects, with an estimated radius of $5.6 \pm0.7$~km from \cite{Chandler:2025} that is comparable to the exocomet radii found for both the $\beta$ Pictoris system and the RZ Psc system, as shown in Figure~\ref{fig:ExocometSizeDistComp}. 1I/Oumuamua and 2I/Borisov both have estimated radii $\le 0.5$~km, making them comparable to the smallest exocomet radii found for the RZ Psc system \citep{Jewitt:2017, Jewitt:2020}. As more interstellar objects are detected with large surveys, we can construct a cumulative size distribution to compare with exocomet systems and the Solar System.  Such comparisons will probe how the same initial population can have considerably different outcomes; the interstellar objects are bodies ejected from their systems, while the exocomets are those that remained.  Furthermore, by building a more complete understanding of small bodies in our stellar neighborhood, we can understand whether interstellar objects like 1I/Oumuamua or 3I/ATLAS are typical of ejected small bodies or outliers.

\begin{figure}
    \centering
    \includegraphics[width=0.99\textwidth]{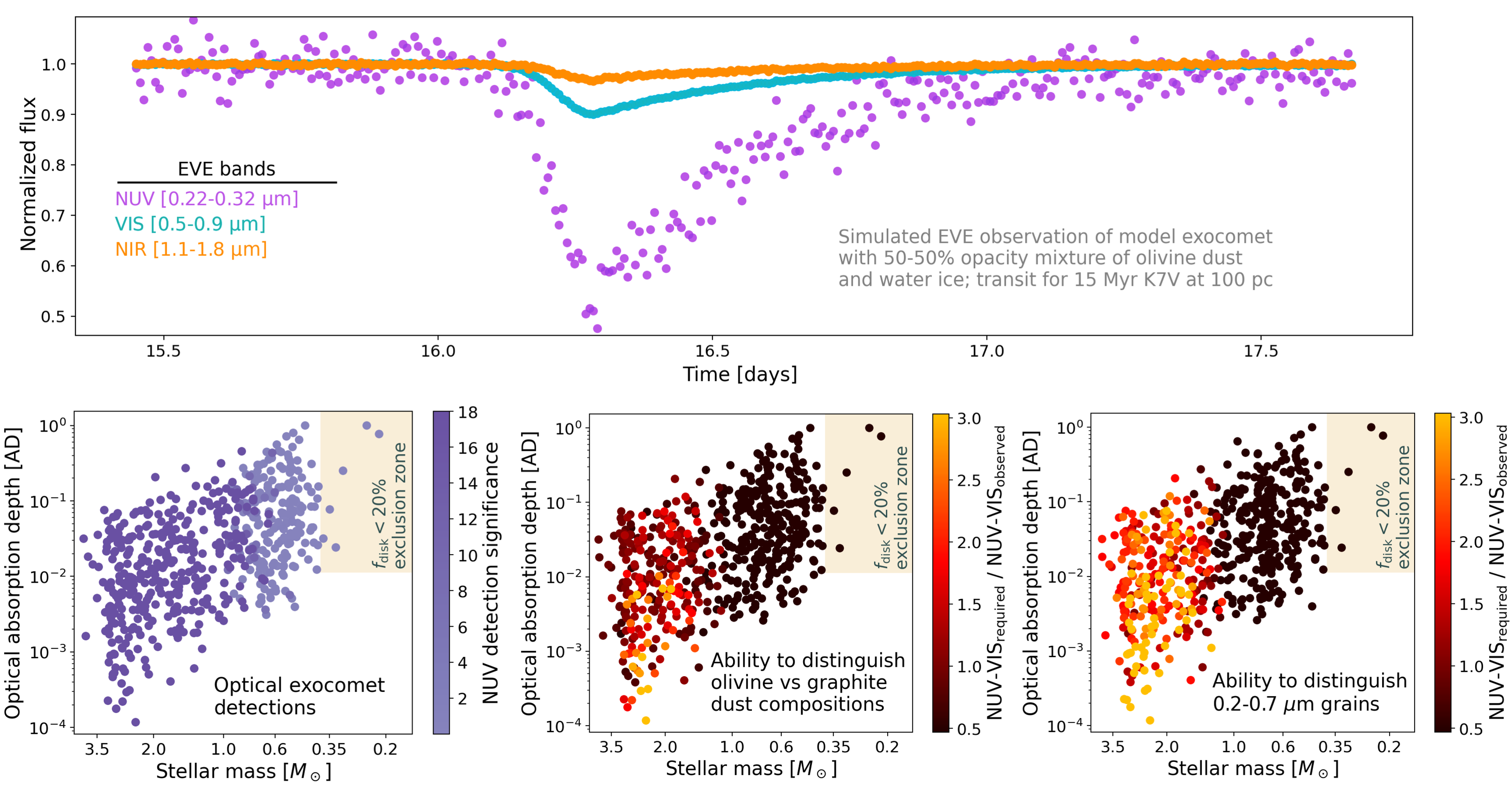}
    \caption{Top: Simulated EVE observation of an exocomet transit of a K7V at 100 pc expected during a 20 d stare at $\sim$20 Myr Theia 43 or 45 associations. We scale the optical light curve template used for the RZ Psc events into the NUV and NIR bands using a 50-50\% weighted average of empirical extinction curves of $\sim$200 nm water ice \citep{kratschmer:1985} and olivine dust \citep{huffman:1975}. Bottom: Distribution of optical ADs versus SpT for 590 simulated detections. The SNR of the NUV detections (left) enable color differences due to composition (middle) and grain size (right) to be distinguished given the color predictions for the narrow-tail multiband comet models of \citet{kalman:2024}. Each panel highlights the region in which EVE should detect exocomets given a 20\% M-star disk fraction, but not for a 2.1\% fraction.}
    \label{fig:eve_yields_maintext}
\end{figure}

\subsection{The Early eVolution Explorer - EVE}\label{sec:EVE}

EVE is a NASA SMEX mission concept under development at the Jet Propulsion Laboratory (JPL).  EVE will observe simultaneously in the near-infrared (NIR, 1100--2000~nm), optical (VIS, 500--900~nm), and near-ultraviolet (NUV, 200--300~nm) with a fast ($<60$~s) cadence, wide field of view (25~deg$^2$), and small pixel scale ($10\arcsec$).  This is the first NASA mission concept focused on the profound early co-evolution of stars and planets and will target at least 17 stellar clusters less than 100~Myr in age.  A precessing low Earth orbit allows EVE to stare continuously at each target cluster from 20--30~days, allowing for long-term monitoring.  The primary science objectives for EVE are to detect dozens of planets younger than 30 Myr to assess their H$_2$O inventory, (2) determine the effects of flares on atmospheric photochemistry, and (3) discover how accretion impacts angular momentum transport in young disks.  However, EVE's unique capabilities enable numerous legacy science investigations across all of astrophysics.

One area in which monitoring of young stellar clusters is likely to have a significant impact is the discovery and multiwavelength-characterization of exocomets. A total of $\sim$70 transiting debris-disk systems will be observed across B7--M4V types, assuming 17 pointings within young clusters for 20~days each and light curves binned to 10~min cadence (e.g., \S \ref{exocomet_yield_code}). If transiting debris disks in the EVE fields have similar exocomet occurrence rates as RZ Psc, which is a particularly collisionally active system, EVE will obtain 5$\sigma$ detections of approximately 120 exocomet transits in the NIR, 590 in the optical, and 560 in the NUV. If transiting debris disks in the EVE fields have similar exocomet occurrence rates as $\beta$ Pictoris, which is a less collisionally active system, EVE will obtain 5$\sigma$ detections of approximately 27 exocomet transits in the NIR, 136 in the optical, and 130 in the NUV. The lower number of NIR detections primarily results from the decreased extinction from water ices and dusty material at longer wavelengths \citep{huffman:1975, kratschmer:1985}. The exocomet yield for 17 potential pointings are shown in Table~\ref{table:EVE}. Unfortunately, neither RZ Psc nor $\beta$ Pictoris are in the potential EVE pointing so follow-up observations of these systems would require an extended mission.  However, that further emphasizes that all of EVE's discovered exocomets will be entirely new detections.

These multi-color detections would enable differentiation of dust grain size and potentially dust compositions. \citet{kalman:2024} simulate exocomet transit light curves due to dust extinction for an A6 star in the Johnson V and R and 2MASS bands across a grid of dust grain sizes and compositions. We interpolate the grid to the EVE optical and NIR bands and extrapolate to the NUV band to ascertain whether EVE NUV-NIR and NUV-VIS colors will be sufficiently sensitive to distinguish between grain sizes of 0.2--0.7~$\mu$m and carbon/graphite vs. olivine dust compositions. The EVE precision curves are binned to 2~hour cadence in order to observe small color changes, the longest cadence for which a $\Delta t$=0.5~day exocomet light curve minimum remains resolved. We compare both the predicted A6V NUV-VIS color difference between 0.2 and 0.7~$\mu$m olivine grains of 180~ppm and predicted A6V NUV-VIS color difference between graphite--olivine dust composition of 110~ppm with the EVE precision curves after scaling the A6V model ADs in each band upwards by the relative $R_*^2$ of the EVE sample. We also compare the other color combinations, although NUV-VIS is the dominant contribution to distinguishing sizes and compositions. This procedure predicts EVE will be capable of distinguishing between 0.2 and 0.7~$\mu$m grain sizes for 50 exocomets at the 3$\sigma$ level. However, EVE will only be able to distinguish between graphite and olivine compositions of $\sim$10 exocomets at the 3$\sigma$ level. 

EVE will also be capable of providing an independent confirmation that the 2.1$^{+4.5}_{-1.7}$\% M-star debris disk fraction observed at long wavelengths is due to detection bias \citep{Luppe:2020} and the true disk fraction is instead comparable to that of FGK stars \citep{Sibthorpe:2018}. Assuming a 20\% disk fraction at 10 Myr, EVE would detect 18$\pm$4 and 38$\pm$6 exocomets from mid- and early M-dwarfs, respectively. If the disk fraction is 2.1$^{+4.5}_{-1.7}$\%, EVE would detect only 3$\pm$1.7 and 4$\pm$2 from mid and early M-dwarfs, respectively. EVE would therefore be able to rule out the lower disk fraction hypothesis at 2.5$\sigma$ and 4.3$\sigma$ confidence levels, respectively. 

\begin{table}
\setlength{\tabcolsep}{3pt}
\centering
\caption{Exocomet Yields for Candidate Young Cluster Pointings}
\begin{tabular}{cccccccccc} % 10 columns now
\hline
\hline
Pointing & R.A. & Dec & Age & $N_\mathrm{RZ\ Psc,\ NUV}$ & $N_\mathrm{RZ\ Psc,\ OPT}$ & $N_\mathrm{RZ\ Psc,\ NIR}$ & $N_\mathrm{\beta\ Pic,\ NUV}$ & $N_\mathrm{\beta\ Pic,\ OPT}$ & $N_\mathrm{\beta\ Pic,\ NIR}$ \\
 & [deg] & [deg] & [Myr] &  &  &  &  &  &  \\
\hline
Theia 13-2 & 83.033 & 0.28 & 10 & 140.8 & 154.93 & 140.8 & 32.63 & 35.9 & 32.63 \\
Theia 13-1 & 83.63 & --4.60 & 10 & 102.02 & 141.43 & 102.02 & 23.64 & 32.77 & 23.64 \\
Theia 133-1 & 51.0 & 48.5 & 55 & 29.06 & 17.17 & 29.06 & 6.73 & 3.98 & 6.73 \\
Theia 45-1 & 242.74 & --22.081 & 15 & 28.7 & 30.16 & 28.7 & 6.73 & 3.98 & 6.73 \\
Theia 77-1 & 59.63 & 34.01 & 8 & 27.32 & 26.49 & 27.32 & 6.33 & 6.14 & 6.33 \\
Theia 17-1 & 57.22 & 33.037 & 5 & 25.63 & 28.58 & 25.63 & 5.94 & 6.62 & 5.94 \\
Theia 22-1 & 122.093 & --47.18 & 12 & 24.99 & 31.93 & 24.99 & 12 & 7.4 & 5.79 \\
Theia 369-1 & 56.8 & 23.96 & 182 & 23.58 & 16.27 & 23.58 & 5.46 & 3.77 & 5.46 \\
Theia 13-3 & 84.82 & 9.70 & 10 & 19.55 & 32.4 & 19.55 & 4.53 & 7.51 & 4.53 \\
Theia 134-1 & 132.78 & --41.94 & 38 & 18.97 & 14.72 & 18.97 & 4.39 & 3.41 & 4.39 \\
Theia 28-1 & 132.78 & --41.94 & 4 & 18.15 & 14.09 & 18.15 & 4.21 & 3.26 & 4.21 \\
Theia 143-1 & 132.78 & --41.94 & 57 & 18.15 & 14.09 & 18.15 & 4.21 & 3.26 & 4.21 \\
Theia 122-1 & 267.0 & --34.5 & 79 & 17.87 & 11.74 & 17.87 & 4.14 & 2.72 & 4.14 \\
Theia 455-1 & 132.78 & --41.94 & 234 & 17.41 & 13.52 & 17.41 & 4.03 & 3.13 & 4.03 \\
Theia 92-2 & 160.004 & --60.18 & 44 & 16.19 & 11.97 & 16.19 & 3.75 & 2.77 & 3.75 \\
Theia 22-2 & 128.80 & --50.51 & 12 & 16.02 & 15.77 & 16.02 & 3.71 & 3.65 & 3.71 \\
Theia 118-1 & 116.07 & --38.48 & 34 & 15.65 & 12.11 & 15.65 & 3.63 & 2.81 & 3.63 \\
\hline
\end{tabular}
\label{table:EVE}
{\newline\newline \textbf{Notes.} Parameters of 17 candidate pointings of 25 deg$^2$ across 96 clusters, given in descending order of NUV exocomet yield. Columns are pointing ID, field center coordinates, age (Zhou et al. 2025), number of NIR exocomets, number of optical exocomets, number of NUV exocomets. Pointings are named by Theia catalog number and subfield designation. We include the yields based on both the RZ Psc exocomet system AD distribution (as the upper limit estimation) and $\beta$ Pictoris system AD distribution (as the lower limit estimation).}
\end{table}

\subsection{Other Future Directions for Exocomet Discovery and Characterization}\label{sec:Missions}
Beyond $\beta$ Pictoris and RZ Psc, five new exocomet hosts have been identified by \citet{norazman:2025} in TESS Sectors 1--26, who estimate an occurrence rate of $\sim$2.6$\times$10$^{-4}$ star$^{-1}$ yr$^{-1}$ with $AD\geq$0.25\%. However, \citet{norazman:2025} exclude candidate transits of $AD>1$\% from their automated search and note incomplete recovery due to heightened stellar activity for the young stars where the disk fractions are highest. These results suggest future searches for young exocomet host stars in the TESS dataset would benefit from joint characterization of stellar and transit signals (e.g., \citealt{Newton:2019, Martioli:2021, mann:2022}). Our distribution suggests automated searches would also benefit from including a wider range of $AD$ values. 

Multi-facility exocomet searches such as that carried out for 5 Vul with TESS and CHEOPS \citep{rebollido:2023} provide another pathway to increase SNR and obtain color information. In particular, high-cadence TESS observations will be complemented in the South by high-precision but low-cadence color information from LSST with the VRO \citep{kalman:2024}. Furthermore, exocomet host stars are likely to overlap the large fields of view of the PLATO long-duration stares \citep{kalman:2024} and Roman Galactic Bulge Time Domain Survey, providing high-cadence, high-precision multi-color and NIR observations, respectively.

Spectroscopic observations could also provide insights into the composition of transiting exocomets.  This has been done for $\beta$ Pictoris using optical and UV absorption spectroscopy which detected various metal lines including Fe II \citep{Ferlet:1987,Kiefer:2014,vrignaud:2024a,vrignaud:2024b,Vrignaud:2025}.  Future observations in the near-infrared could search for volatile species such as CO$_2$, CO, and H$_2$O.  These species are most commonly detected in Solar System comets and have been detected recently from 3I/ATLAS with JWST \citep{cordiner:2025}.

\section{Conclusions}\label{sec:conclusion}

In this paper, we present an analysis of three sectors of red-optical bandpass TESS observations of the young debris disk system RZ Psc, with cadences of 30 minutes and 200 seconds; we summarize the main takeaways:
\begin{enumerate}
    \item We identify 24 exocomets transiting RZ Psc.  These observation suggest an exocomet transit rate of 0.40 exocomets day\textsuperscript{-1}, about double what has been reported for the $\beta$ Pictoris system at $\sim$0.19 exocomets day\textsuperscript{-1} \citep{lecavelier:2022}.
    \item In Section~\ref{sec:sizedistcomp}, we present an exocomet size distribution and report radii between $0.77 \pm0.17$~km and $7.31 \pm 0.33$~km. We compare the exocomet size distribution of the RZ Psc system to size distributions of populations in the Solar system and the $\beta$ Pictoris system and determine that our size distribution is consistent with the slope of several Solar system size distributions and contains a break predicted by two KBO size distribution models.
    \item In Section~\ref{sec:EVE}, we determine that the EVE NASA SMEX mission concept will be capable of detecting significant populations of exocomets and will enhance our understanding of small bodies in young systems. EVE will obtain 5$\sigma$ detections of approximately 120 exocomet transits in the NIR, 590 in the optical, and 560 in the NUV, which will not only allow for further exocomet system characterization but will also enable the dust grain size and potentially compositions to be distinguished.
\end{enumerate}

\section{Acknowledgements}

M.A.M. acknowledges support for part of this research from the National Aeronautics and Space Administration (NASA) under award number 19-ICAR19\_2-0041.  
W.H. received funding through the NASA Hubble Fellowship grant HST-HF2-51531 awarded by STScI, which is operated by the Association of Universities for Research in Astronomy, Inc., for NASA, under contract NAS5-26555. A.M.C. was supported in part by NASA under grant No. 80NSSC21K0398 issued through the NNH20ZDA001N Astrophysics Exoplanets Research Program (XRP). Part of this research was carried out at the Jet Propulsion Laboratory, California Institute of Technology, under contract with the National Aeronautics and Space Administration.
The TESS data presented in this article were obtained from the Mikulski Archive for Space Telescopes (MAST) at the Space Telescope Science Institute. The specific observations analyzed can be accessed via \href{https://doi.org/10.17909/1xdh-2r79}{doi: 10.17909/1xdh-2r79}.

\software{\texttt{astropy} \citep{astropy:2013,astropy:2018,astropy:2022} 
\texttt{eleanor} \citep{Feinstein:2019, Brasseur:2019}}

\clearpage
\appendix
\setcounter{figure}{0}
\renewcommand{\thefigure}{A\arabic{figure}}
\setcounter{table}{0}
\renewcommand{\thetable}{A\arabic{table}}
\section{Stellar Variability Model}\label{GP_model}

\begin{figure}
    \centering
    \includegraphics[width=0.8\textwidth]{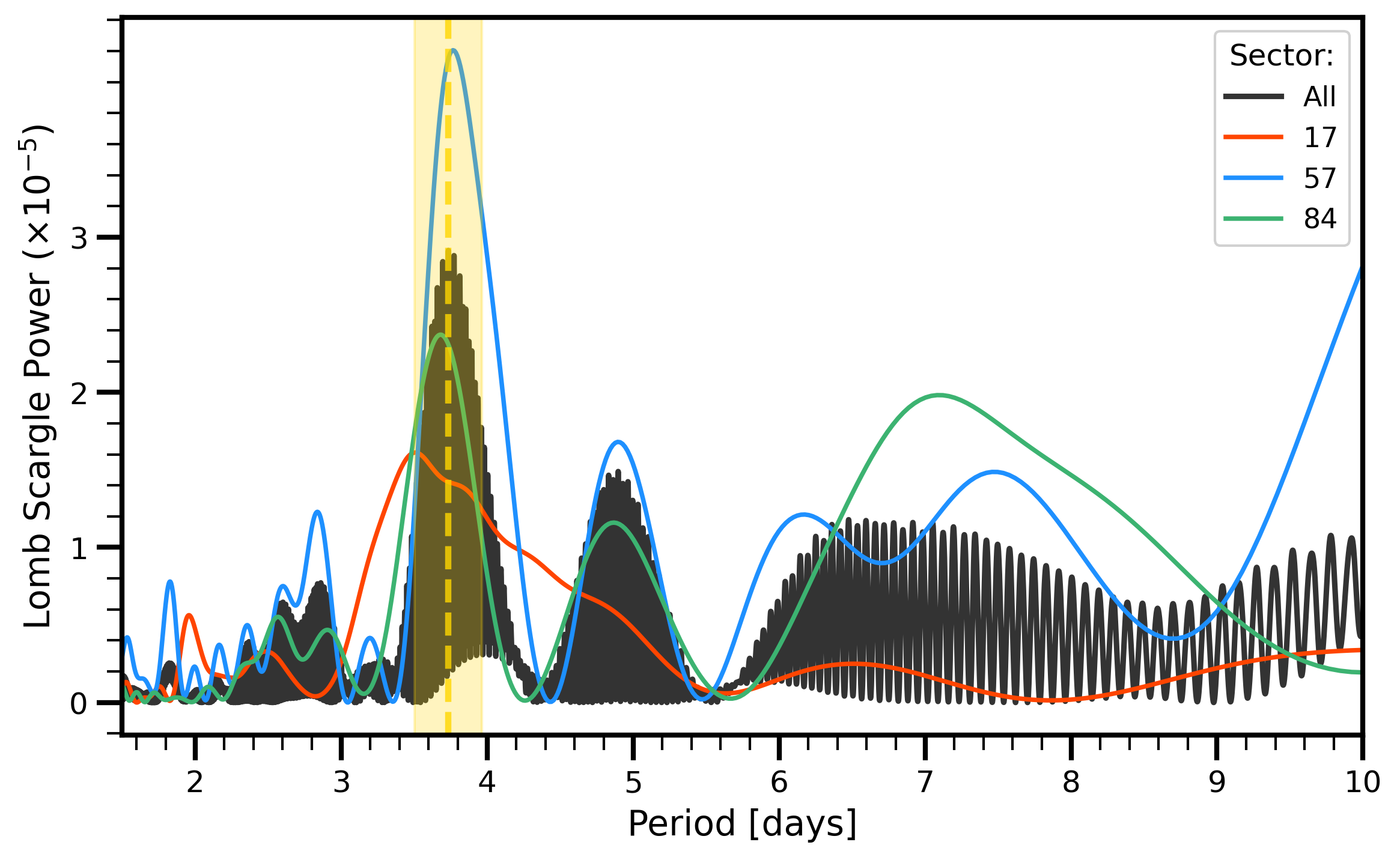}
    \caption{The Lomb-Scargle periodogram, computed from the combined TESS data across all three sectors (black), with the initial exocomet models subtracted, shows that the maximum period of 3.73$\pm$0.2~d is consistent across all sectors and is indicated here by the dashed yellow line, with the 1$\sigma$ uncertainty shaded. Some of the later peaks are harmonics of the 3.73-day period. The Lomb-Scargle periodogram for the individual sectors is overplotted: sector 17 (red), sector 57 (blue), and sector 84 (green).}
    \label{fig:LombScargle}
\end{figure}

We compute a Lomb-Scargle periodogram implemented with \texttt{exoplanet.estimators.lomb\_scargle\_estimator}\footnote{\url{https://github.com/exoplanet-dev/exoplanet}} shown in Figure~\ref{fig:LombScargle} and find a maximum period of 3.73$\pm$0.2~days; this is consistent with the rotation period of $\sim3.6$~days computed by \cite{potravnov:2017}. We fit initial exocomet models to the transits identified using the model described in Section~\ref{sec:modeling}. Subtracting these initial fits allowed us to create a Gaussian Process (GP) model for stellar variability. The total light curve duration of the GP fit is longer than the exocomet transit durations; the in-transit times are excluded from the GP fits entirely, and only the stellar signals are fit.  We then subtracted the stellar variability model from the initial light curve, leaving only the exocomet transits. We then refit these exocomet transits to obtain the final exocomet parameters, as described further in Section~\ref{sec:sizedist}. 

\begin{figure}
    \centering
    \includegraphics[width=\textwidth]{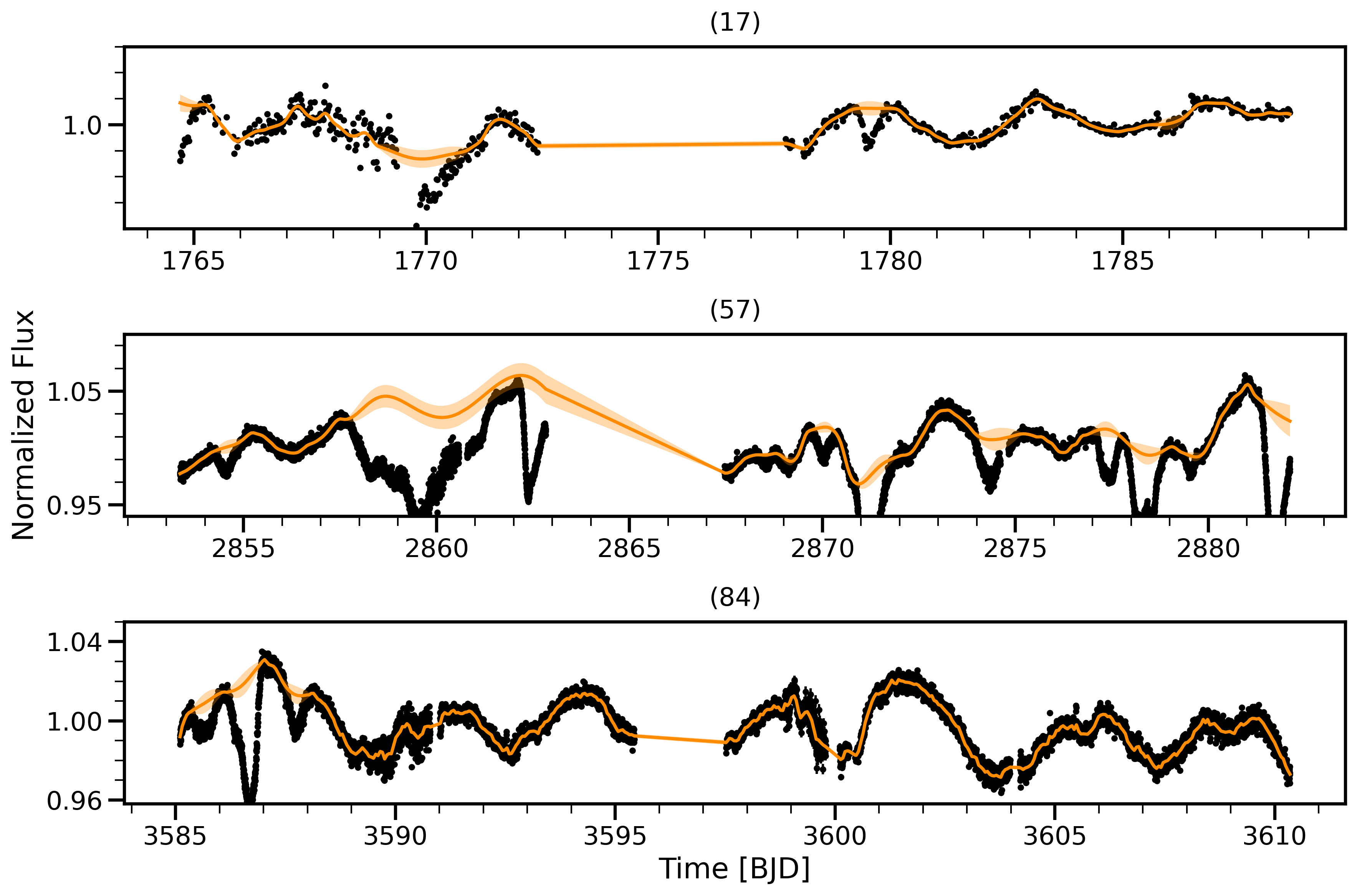}
    \caption{GP models for all sectors of RZ Psc, the center value is shown by the solid orange line, with the 1$\sigma$ uncertainty shown by the orange shaded region. This uncertainty was propagated into the exocomet model parameters, from which radii and transit times were calculated in Section~\ref{sec:sizedist}.}
    \label{fig:GPmodelfig}
\end{figure}

The stellar variability GP model was the sum of two GP kernels: one for rotational variability and one for non-rotational variability, as well as a term for excess jitter the rotational variability and non-rotational variability terms may not account for. The GP model fits the light curves of RZ Psc, which are shown in Figure~\ref{fig:GPmodelfig}. The \texttt{pymc3}\footnote{\url{https://github.com/aseyboldt/pymc3}} and \texttt{exoplanet} packages were used to execute this model. Table~\ref{table:Bfield_tab} lists the GP model parameters used to calculate the GP model period and $1\sigma$ uncertainty. A detailed description of the GP model used here can be found in Section 6 of \cite{Howard:2021}.

\begin{table}[h!]
\centering
\caption{Stellar rotation GP hyper-parameters}
\begin{tabular}{cccc}
\hline
Hyper-parameter & Prior & Value $\mu$ & Bounds \\
\hline
\texttt{log\_amp} & Gaussian & $\log(\mathrm{Var}(\Delta T))$ & $\sigma$=5.0 \\
\texttt{log\_period} & Gaussian & $\log(P_{\mathrm{rot}})$ & (0.0, $\log$ 50.0)) \\
\texttt{log\_Q0} & Gaussian & 1.0 & $\sigma$=10.0 \\
\texttt{log\_deltaQ} & Gaussian & 2.0 & $>$0, $\sigma$=10.0 \\
\texttt{mix} & Uniform & --- & (0.0––1.0) \\
\texttt{logSw4} & Gaussian & $\log(\mathrm{vvar}(\Delta T))$ & $\sigma$=5.0 \\
\texttt{logw0} & Gaussian & $\log(2\pi/10)$ & $\sigma=5.0$ \\
\texttt{logs2} & Gaussian & $2\log(\mathrm{vvar}(\Delta T_{\mathrm{err}}))$ & $\sigma$=2.0 \\
\texttt{mean} & Gaussian & $0.0$ & $\sigma$=1.0 \\
\hline
\end{tabular}
\label{table:Bfield_tab}
\end{table}

\section{Exocomet Radii Calculation}\label{exocomet_radii_calculation}
Here we include the equations used for the radii calculation, shown in Figure~\ref{fig:ExocometSizeDistComp}, using the absorption depth and the transit time:

\begin{equation}\label{eqn:AD}
    AD \simeq 5\times10^{-5} (\frac{\dot{M}_{1 \ au}}{10^5\ kg\ s^{-1}})(\frac{q}{1\ au})^{-1/2}(\frac{M_*}{M_\odot})
\end{equation}

\begin{equation}\label{eqn:q}
    q = (4GM_*{\Delta}t^2)/(\pi^2R_*^2)
\end{equation}

where $q$ is the orbital periastron distance, ${\Delta}$t is the transit time, $\dot{M}_{1 \text{ au}}$ is the dust production rate of the comet when it is at 1 au from the star defined in \cite{lecavelier:2022} as:

\begin{equation}\label{eqn:Mdot}
    \dot{M}_{1\text{ au}} \simeq 3.5\times10^{10}\text{ kg s}^{-1}\ \left(\frac{R}{30\text{ km}}\right)^2
\end{equation}

\noindent where $R$ is the radius of the exocomet. This equation was calculated by \cite{lecavelier:2022} by scaling the observation of the Hale-Bopp comet. We replaced the Hale-Bopp dust production rate with a mass loss rate for RZ Psc from \cite{su:2023}, and removed the luminosity correction factor. We used the mass loss rate from \cite{su:2023} because the RZ Psc system is more collisionally active than the Solar system and the $\beta$ Pictoris system. This is evidenced in the higher exocomet transit rate for RZ Psc, and the system should not be assumed to have the same mass loss rate. By using the stellar parameters of RZ Psc, stated in Section~\ref{sec:intro}, we derived an equation for the exocomet radius using $\Delta$t and $AD$.

In addition to the first approach, we employ a second independent method of calculating the exocomet radii from the AD. Specifically, by observing a break in the AD distribution, shown in Figure~\ref{fig:ExocometSizeDistComp}, we can use equations 6 and 7 from \cite{Pan:2005} to calculate $r_\text{break}$ based on the age of the system \citep[$20^{+3}_{-5}$~Myr, $30-50$~Myr][]{potravnov:2019, punzi:2018}, given the two varying estimations of age we use an average age between these two estimates of 35~Myr. We use $\beta=3$, assuming momentum conservation; reasons for this are discussed in Section~\ref{sec:disc}; this yields an $r_\text{break} \simeq  2.26$~km.  

We apply the following relation to determine a factor for converting absorption depths to radii:

\begin{equation}\label{eqn:AD_to_rad}
    f = \frac{r_\text{break}}{AD_\text{break}}
\end{equation}

We find that the two methods described here yield a highly similar size distribution, shown in Figure~\ref{fig:ExocometSizeDistComp}. The mean difference in radii is $ \sim0.5$~km.

\begin{figure}
    \centering
    \includegraphics[width=0.99\textwidth]{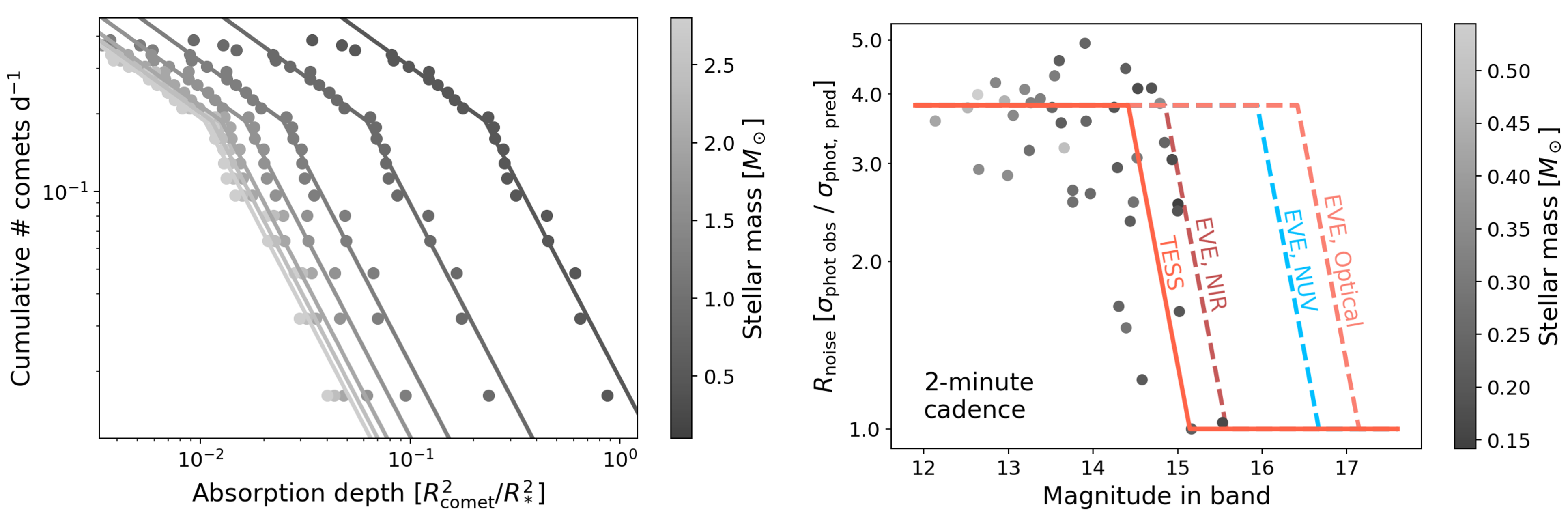}
    \caption{Left: Dependence of the pipeline scaling of the measured RZ Psc cumulative occurrence rate by stellar mass Right: Additional noise from stellar activity observed in 2 min cadence TESS light curves of candidate EVE sources relative to that expected from the TESS precision curve. The magnitude at which the decrease in TESS precision overtakes the relatively constant stellar activity signal is predicted for the EVE bands and used to construct stellar activity precision-vs-magnitude relations.}
    \label{fig:eve_yields_suppl}
\end{figure}

\section{Exocomet Yield Prediction Details with the Early eVolution Explorer}\label{exocomet_yield_code}
Several dozen 25~deg$^2$ EVE candidate fields were identified within the dense core regions of nearby young associations and moving groups using a grid search to maximize early exoplanet, stellar flare, and accretor yields (\citealt{howard:2025}, Burt et al., in prep. Venuti et al., in prep.). We predict exocomet transit event yields for the top 32 fields, at least 17 of which will be observed for $\sim$20~day each. First, the estimated photometric precision-versus-magnitude curves of each band are binned from a 30--60~s base cadence to 10~min cadence in order to both resolve and maximize signal from exocomet light curves. The precision curves, TESS magnitude, and estimated stellar SED of each source in the field of view are used to derive the smallest exocomet absorption depth ($AD$) detectable at 5$\sigma$ on a per-star basis. NUV extinction factors are applied to the NUV precision curves as described in \citet{howard:2025}. The photometric precision curve of the TESS facility from \citet{stassun:2018} is also included for yield validation on similarly bright sources to RZ Psc.

Next, the RZ Psc exocomet cumulative frequency distribution is scaled by $R_*^2/R_\mathrm{RZ~Psc}^2$ to account for the differing stellar contrast of each source to obtain the source exocomet rate for each band in the presence of a transiting disk, $\nu_{\mathrm{AD_{band}},{5\sigma}}(R_\mathrm{*}, m_\mathrm{*})$, as shown in Fig. \ref{fig:eve_yields_suppl}. Here, $\nu_{\mathrm{AD_{band}},{5\sigma}}(R_\mathrm{n}, m_\mathrm{n})$ is a broken power-law function for the RZ Psc exocomet rate with separate fit parameters above and below the AD value at which a power-law break exists (Eq. \ref{eqn:brokenpowerAD}), and where $R_\mathrm{*}$ and $m_\mathrm{*}$ are the stellar radius and band magnitude, respectively:
\begin{equation}
\label{eqn:brokenpowerAD}
    \log\nu_{\mathrm{AD_{band}},{5\sigma}}(R_\mathrm{*}, m_\mathrm{*})=
    \begin{cases}
        -0.59\log_{10}(AD)-0.88\log_{10}(R_*^2/R_\mathrm{RZ~Psc}^2), & AD < AD_\mathrm{break} \\
        -1.56\log_{10}(AD)-1.15\log_{10}(R_*^2/R_\mathrm{RZ~Psc}^2), & AD \geq AD_\mathrm{break}
    \end{cases}
\end{equation}
We compare the observed exocomet rate of RZ Psc against the Eq. \ref{eqn:brokenpowerAD} prediction for the TESS facility and band for sources of similar brightness and spectral type. This comparison reveals the observed rate is consistently $\approx$4$\times$ lower than expected from the TESS precision curve, which is suggestive of stellar noise. We test the stellar origin hypothesis by comparing the observed ($\sigma_{phot,obs}$) and predicted ($\sigma_\mathrm{phot,~pred}$) precision using the sample of 2-min cadence TESS observations of 41 young sources located within EVE candidate pointings as described in \S 6.2 of \citet{howard:2025}. Large flares are removed and the remaining data binned to cadences from 4--60 min and searched by eye to identify coherent variability signals, which appear clearest at 30 min cadence. These signals appear consistent with stellar activity (e.g. low-level flaring). The amplitude of the residual variability is measured as the absolute difference between the maximum and minimum flux in each time bin, after removing outliers with a 3$\sigma$ clip. The same process is repeated for a simulated 2-min cadence light curve with Gaussian noise of $\sigma_\mathrm{phot,pred}$ given by the TESS precision-vs-magnitude curve. The $\mathcal{R}_\mathrm{noise}=\sigma_\mathrm{phot,obs}$/$\sigma_\mathrm{phot,pred}$ values of each source are then recorded. Fig. \ref{fig:eve_yields_suppl} shows the resulting relationship between $\mathcal{R}_{noise}$ versus TESS mag and stellar mass. The observed scatter is 3.8$\times$ worse at $T<$14.4, follows a log-linear decrease from 14.4$<T<$15.1, and converges with $\sigma_\mathrm{phot,pred}$ for $T>$15.1. This behavior is consistent with a stellar activity jitter noise floor in the TESS precision curve until the $\sigma_\mathrm{phot,~pred}$ amplitude exceeds the jitter amplitude at fainter magnitudes. Correcting the TESS precision curve by the $\mathcal{R}_\mathrm{noise}$ curve brings the predicted exocomet yields into alignment with the observed rate of RZ Psc. We therefore scale the $\mathcal{R}_\mathrm{noise}$ curve to the equivalent magnitudes in the EVE bands for the SNR values of the log-linear decrease. The resulting $\mathcal{R}_\mathrm{noise}$ curves are shown in Fig. \ref{fig:eve_yields_suppl} and used to correct the EVE precision curves.

We then estimate the debris disk fraction in each field using the \citet{wyatt:2008} disk-fraction-versus-age relations for BA and FGK stars. The M star relation is less certain, but is obtained by scaling the FGK relation to give a 20\% disk fraction at 10 Myr in agreement with values inferred from \textit{Herschel} observations \citep{Luppe:2020}. Likewise, we compute geometric transit probabilities from the stellar radii and orbital distances, inferred from the observed exocomet transit durations $\Delta t_\mathrm{comet}$ using Eq. 6 of \citet{heller:2024}. We obtain semi-major axes of $a_{OBAFG}$=0.5--0.7, $a_{K}$=0.3--0.6 au, and $a_{M}$=0.3--0.9 au from $\Delta t_\mathrm{comet}$=0.5, 0.35, and 0.25~day for the OBAFG, K, and M star populations respectively. The KM star values are reduced from $\Delta t_\mathrm{comet}$=0.5~day to match the observed transit probabilities of typical low-mass systems \citep{dressing:2015}. Finally, the disk fractions, transit probabilities, cumulative event rates, and stare time are combined to estimate exocomet yields. The total number of comet detections per field $N_\mathrm{field}$ per band is obtained by summing over the yields of the individual stars in each field of view (FoV) according to Eq. \ref{eqn:comet_yields}:
\begin{equation}
\label{eqn:comet_yields}
    N_\mathrm{FoV} = \sum_{n=1}^{n_\mathrm{stars}(FoV)} \nu_{\mathrm{AD},{5\sigma}}(R_\mathrm{n}, m_\mathrm{n}) \times t_\mathrm{obs}
\end{equation}
Here, $\nu_{\mathrm{AD},{5\sigma}}$ is the rate at which transits occur with an amplitude exceeding a 5$\sigma$ detection for a given SNR curve, $R_\mathrm{*,n}$ is the stellar radius, and $m_\mathrm{*,n}$ is the stellar magnitude in the band. The yields obtained for all 32 separate pointings are then sorted in descending order and the top 17 pointings are added together to produce a minimum total exocomet yield over the mission lifetime, $N_\mathrm{comet}$ = $\sum_{p}^{17} N_\mathrm{pointing}(p)$. In total, we find EVE should detect 580 exocomets at 10~min cadence given 17 pointings of 20~day each, with at least 120 comets detected simultaneously in all three NUV, optical, and NIR bands.

\clearpage
\bibliography{References.bib}
%\bibliography{sample631}{}
%\bibliographystyle{aasjournal}
%\bibliographystyle{aasjournal}
%\bibliography{References}

\end{document}